\documentclass{mem}
\usepackage{natbib}
\usepackage{txfonts}
\usepackage{balance}
\usepackage{graphicx}
\usepackage{txfonts} 
\usepackage[a4paper]{hyperref}
\idline{1}{1}
\begin{document}

\title{Simulating the WFXT sky}

\subtitle{}

\author{P. \, Tozzi\inst{1}\inst{,2}, J. Santos\inst{1},
  H. Yu\inst{1}, A. Bignamini\inst{3}, P. Rosati\inst{4},
  S. Borgani\inst{2}\inst{,3}, S. Campana\inst{5}, P. Conconi\inst{5},
  R. Gilli\inst{6}, M. Paolillo\inst{7}, A. Ptak\inst{8}, \and the WFXT Team}

\offprints{P. Tozzi}
 
\institute{
INAF -- Osservatorio Astronomico di Trieste, Via Tiepolo 11, I-34131 Trieste, Italy \email{tozzi@oats.inaf.it}
\and
INFN, Sezione di Trieste, via Valerio 2, I-34127 Trieste, Italy
\and
Dipartimento di Fisica, Sezione di Astronomia, Universit\`a di Trieste, via Tiepolo 11, I-34143 Trieste, Italy
\and
ESO - European Southern Observatory, D-85748 Garching bei Munchen, Germany 
\and
INAF - Osservatorio Astronomico di Brera, via via Brera 28, 20121 Milano, Italy
\and
INAF - Osservatorio Astronomico di Bologna, via Ranzani 1, I-40127 Bologna, Italy
\and
Universit\`a Federico II, Dip. di Scienze Fisiche, via Cintia, I-80126, Napoli
\and
NASA Goddard Space Flight Center, Maryland, USA}

\authorrunning{P. Tozzi \& the WFXT Team}

\titlerunning{Simulating the WFXT sky}

\abstract{We investigate the scientific impact of the Wide Field X-ray
  Telescope mission.  In order to quantify the size and the properties
  of the WFXT data sets, we present simulated images and spectra of
  X--ray sources as observed from the three surveys planned for the
  nominal 5-year WFXT lifetime.  The goal of these simulations is to
  provide WFXT images of the extragalactic sky in different energy
  bands based on accurate description of AGN populations, normal and
  star forming galaxies, groups and clusters of galaxies.  The images
  are realized using a detailed PSF model, instrumental and physical
  backgrounds/foregrounds, accurate model of the effective area and
  the related vignetting effect.  Thanks to this comprehensive
  modelization of the WFXT properties, the simulated images can be
  used to evaluate the flux limits for detection of point and extended
  sources, the effect of source confusion at very faint fluxes, and in
  general the efficiency of detection algorithms.  We also simulate
  the spectra of the detected sources, in order to address specific
  science topics which are unique to WFXT.  Among them, we focus on
  the characterization of the Intra Cluster Medium (ICM) of high-z
  clusters, and in particular on the measurement of the redshift from
  the ICM spectrum in order to build a cosmological sample of galaxy
  clusters.  The end-to-end simulation procedure presented here, is a
  valuable tool in optimizing the mission design.  Therefore, these
  simulations can be used to reliably characterize the WFXT discovery
  space and to verify the connection between mission requirements and
  scientific goals.  Thanks to this effort, we can conclude on firm
  basis that an X-ray mission optimized for surveys like WFXT is
  necessary to bring X-ray astronomy at the level of the optical, IR,
  submm and radio wavebands as foreseen in the coming decade.

 }

\maketitle{}

\section{Introduction}

The strong interest behind the Wide Field X--ray Telescope (WFXT)
stems from the fact that no planned or foreseen X-ray mission is
optimally designed for surveys.  The tremendous scientific impact that
we experienced in the last ten years thanks to the Chandra and
XMM-Newton X-ray telescopes, largely relies on the previous all--sky
surveys performed by satellites like Einstein or ROSAT, which were
able to provide a large number of potentially interesting X-ray
targets.  A new, deeper, wide-angle X-ray survey is needed in order to
perform a significant step forward in the field of X-ray astronomy.
In addition, the innovative concept of WFXT
\citep{mur08}\footnote{http://www.wfxt.eu; http://wfxt.pha.jhu.edu}
will allow one not only to deliver catalogs of X-ray sources, but also
to characterize most of them, and achieve several scientific goals
well in advance of a multiwavelength follow-up.  Eventually, the
synergies with future surveys in other wavebands will greatly enhance
its scientific impact (see Rosati et al. this volume).  In our view,
the WFXT mission will provide an immense legacy value for Galactic and
extragalactic astronomy.

Simulations of data products from the planned WFXT surveys are crucial
in order to quantify the scientific impact of the mission.  In order
to achieve this goal, we set up a procedure to build images of
extragalactic fields and spectra of the sources according to the
instrument design of WFXT.  We put a strong effort in the modelization
of the X-ray source populations, in order to investigate in detail
several science cases which are unique to WFXT.  Among these
scientific cases we discuss the study of the ICM in groups and
clusters of galaxies, the building of a cosmological sample of
clusters whose redshift is obtained directly by the X-ray spectral
analysis, and the characterization of the different classes of AGN up
to redshift $z\sim 6$ \citep{giac09,mur09,vik09}.  Many other science
cases can be addressed with these simulations, including cases
relevant for Galactic astronomy and the local Universe \citep{ptak09},
but are not discussed here.

This paper is organized as follows.  In \S 2 we derive the main
quantities which are relevant to the simulations like the background
and foreground components and the typical conversion factors, and show
how we build a mock WFXT image.  In \S 3 we describe the modelization
of the extragalactic source populations adopted in our simulations.
In \S 4 we describe the results of a preliminary analysis using a
simple detection algorithm to provide a conservative estimate of the
capability of WFXT of detecting X-ray sources.  In \S 5 we describe
the spectral simulations and focus on the capability of measuring the
redshift for a large sample of groups and clusters of galaxies.
Finally our conclusions are summarized in \S 6.

\section{Building WFXT simulated images}

\begin{figure}[]
\resizebox{\hsize}{!}{\includegraphics[clip=true]{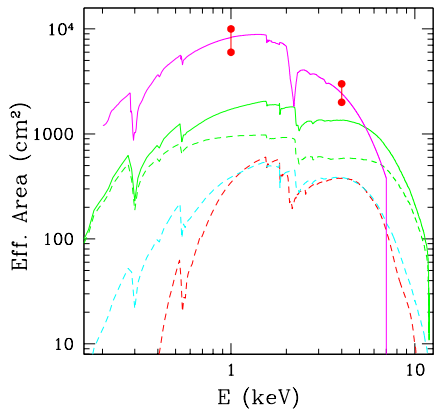}}
\caption{ \footnotesize Effective area for WFXT design (magenta) at
  the aimpoint for the sum of the three modules.  Red dots show the
  requirement and the goal at 1 and 4 keV.  The {\sl
    Chandra} effective area is shown as a red dashed line, while the
  total XMM-Newton response is shown as a solid green line (green
  dashed for PN and cyan dashed for MOS).  }
\label{eff_area}
\end{figure}

The details of the WFXT design are presented in the contribution by
Pareschi \& Campana in this volume.  Here we recall the most relevant
properties which affect our simulations.  First we consider the
sensitivity of the X-ray telescope.  The total effective area of the
three modules of the WFXT at the aimpoint is shown in Figure
\ref{eff_area}.  In the soft band, the effective area of WFXT is about
one order of magnitude larger than the total effective area of XMM
(the sum of the PN and the two MOS), and about two orders of
magnitudes larger than the Chandra one.  The effective area in the
hard band is comparable to the total value of XMM at 5 keV.  Achieving
a large effective area in the soft band while still keeping a
significant area in the hard band (2-7 keV) is key to characterize the
detected sources.

\begin{figure}[]
\resizebox{\hsize}{!}{\includegraphics[clip=true]{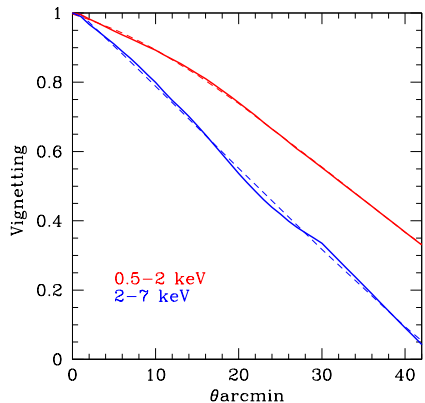}}
\caption{ \footnotesize Vignetting with respect to the aimpoint as a
  function of the off-axis angle at 1 keV (red solid line) and 4 keV
  (blue solid line).  Dashed lines show the analytical fits used in
  the simulations.}
\label{vign}
\end{figure}

With its very large field of view (1 square degree), WFXT images
suffer an important vignetting effect, which consists in a decrease of
the effective area with the off--axis angle with respect to the
aimpoint (assumed to be in the center of the image). This is an
important aspect to be included in the simulations, since it allows us
to compute correctly the number of detected photons for the sources
randomly distributed in the field. The vignetting in both bands for a
single pointing is shown in Figure \ref{vign}.

\begin{figure}[]
\resizebox{\hsize}{!}{\includegraphics[clip=true]{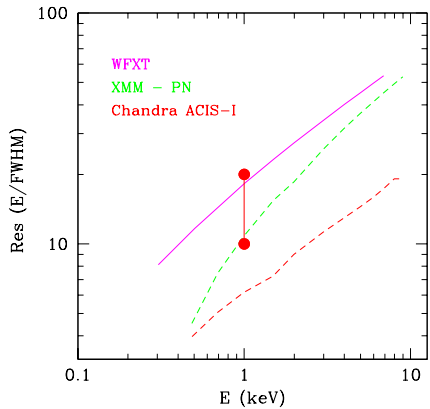}}
\caption{ \footnotesize Spectral resolution $E/\Delta E$ for WFXT
  (magenta solid line). Goal and requirement are shown as red
  dots. Spectral resolution for the XMM PN detector is shown in green,
  while the Chandra ACIS-I in red. }
\label{specres}
\end{figure}

The spectral resolution, defined as $ E / \Delta E$, where $\Delta
E$ is the Half Energy Width (HEW) of an emission line with zero
intrinsic width, is shown in Figure \ref{specres}.  Spectral
resolution is about a factor of 3 better than the Chandra (ACIS-I)
resolution over the whole energy band, while it is comparable to the
XMM resolution (PN) above 1 keV. Spectral resolution is crucial to
perform line diagnostic. As we will see, the ability to measure the
ubiquitous Fe $K_\alpha$ line in the ICM spectra is one of the key
requirements in order to build a cosmological sample of groups and
clusters of galaxies without recurring to time--expensive optical
follow-up (see \S 5).

\begin{table}
\label{cf_sim_table} 
\centering                          
\begin{tabular}{rrrrr}
\hline \hline \\
Source  & $0.5-2$ keV & $2-7$ keV  \\
\\
\hline \hline \\
XRB, $\Gamma = 1.4$  & $ 2.25 \times  10^{-13}$  &  $2.35 \times  10^{-12}$   \\
AGN z=1, $N_H = 10^{21}$ & $ 2.19\times  10^{-13}$  &  $2.06 \times  10^{-12}$  \\
AGN z=1, $N_H = 10^{23}$  & $ 3.45\times  10^{-13} $ &  $2.40  \times  10^{-12}$  \\
SF Gal $\Gamma = 1.9$  & $ 2.16 \times  10^{-13}$ &  $2.05 \times  10^{-12}$  \\
ICM $z=0.5, kT=5$  & $2.22 \times  10^{-13}$  & $1.85 \times  10^{-12}$  \\
\\
\hline \hline \\
\end{tabular}
\caption{Typical Energy Conversion Factors (ECF) at the aimpoint.  The
  units are erg s$^{-1}$ cm$^{-2}$/(cts s$^{-1}$).  We assume a
  typical Galactic absorption of $N_H = 3 \times 10^{20}$ cm$^{-2}$. }
\end{table}

An important figure which is obtained directly from the WFXT spectral
response, is the Energy Conversion Factor (ECF) appropriate to a given
spectral shape (i.e., a given source class).  The ECFs are defined as
the ratio of the observed energy flux in a given band and the observed
photon rate in the same band, and are computed always at the aimpoint
(so they must be weighted by the vignetting whenever the observed
source is not at the aimpoint).  In Table \ref{cf_sim_table} we show
the conversion factors for some typical extragalactic source in the
soft and hard bands.  
These values
are posted here as a reference: in the simulations the net photon rate
of each source is computed exactly according to its spectral shape.

The next aspect we consider is the noise, which is due to the sum of
several components. One is the particle (or instrumental) background,
while the other contributions are from astrophysical sources. The
galactic foreground is truly diffuse, with fluctuations of the order
of a few percent.  The extragalactic unresolved background is given by
undetected point sources and it depends on the flux limit at which the
extragalactic sources can be resolved, and therefore it is also a
function of the exposure time.  In our simulator we include the
unresolved AGN contribution and the unresolved ICM contribution as a
uniform distribution spread across the FOV.  As a useful reference we
show the expected average values in Table \ref{bck_tot} in photons per
second per field of view (one square degree).  The particle background
is very low thanks to the low Earth orbit proposed for WFXT.  The
Galactic background is by far the dominant component in the soft band
(actually below 1 keV), while the extragalactic background, due to
point sources (mostly AGN) and groups and clusters, strongly depends
on the different exposure of the Wide, Medium and Deep surveys, given
the different minimum flux at which it is possible to detect single
sources.

\begin{table}
\centering                          
\begin{tabular}{cccc}
 \hline\hline\\
 Source & $0.5-2$  keV & $2-7$ keV\\
\hline\hline\\
Particles &      0.188   & 0.397 \\ 
Galactic &  21.4       &  0.0 \\ 
AGN wide  &   9.5      &  3.13 \\ 
AGN medium &     3.9    &  1.65 \\ 
AGN deep &      0.8   & 0.17 \\ 
Cluster wide &  1.46  &  0.3 \\ 
Cluster medium &  0.79    &  0.14 \\ 
Cluster deep &  0.2  & 0.03 \\ 
 \hline\hline \\
\end{tabular}
\caption{Total background photon rates (cts/s) in the soft and hard
  bands for one FOV (1 deg$^2$)}
\label{bck_tot}
\end{table}

The information given so far can be used to estimate the collected
photons from several extragalactic sources in each of the three WFXT
planned surveys.  However, a proper estimate of the detectability, or
the signal-to-noise ratio, of each source, depends also on the PSF.
Thanks to the polynomial X-ray mirrors \citep{bbg92} the PSF has an HEW
of 5 arcsec almost constant across the entire field of view.  This is
the key property which makes WFXT the ultimate X-ray survey mission,
since it allows one to detect and characterize pointlike and extended
sources without hitting source--confusion down to very low fluxes.
Together with the hard band sensitivity, the spatial resolution
constitute the main difference of WFXT with respect to other planned
X-ray missions like eRosita \citep{pr09}.

\begin{figure}[]
\resizebox{\hsize}{!}{\includegraphics[clip=true]{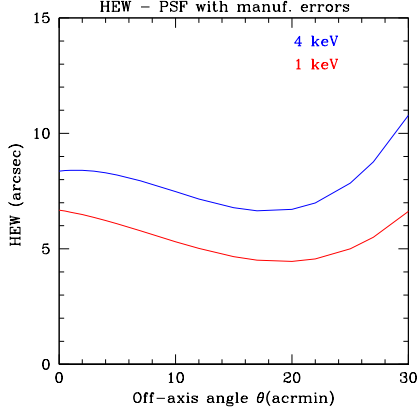}}
\caption{ \footnotesize Half Energy Width of the PSF (including
  estimated manufacturing errors) in the soft (1 keV, red line) and
  hard band (4 keV, blue line) as a function of the off--axis angle
  $\theta$. }
\label{psf}
\end{figure}

A realistic image simulator needs a detailed modelling of the PSF.
The HEW of the PSF is shown in Figure \ref{psf}, at two different
energies, 1 and 4 keV, representative of the soft and hard bands.
Note that the images are realized taking into account all the features
of the PSF, including its asymmetric shape \citep{conc}.


This short review of the WFXT instrumental properties is sufficient to
comprehend the main ingredients of the WFXT mock images we realized.
In order to achieve a realistic rendition of the X-ray sky, we now
exploit what we have learned from deep X-ray extragalactic surveys to
date on different source populations.

\section{Source populations in extragalactic fields}

\begin{figure}[]
\resizebox{\hsize}{!}{\includegraphics[clip=true]{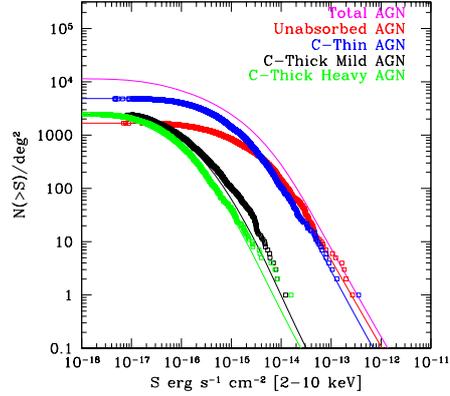}}
\caption{ \footnotesize The hard-band logNlogS from the mock input
  catalog of extragalactic point sources (empty squares) compared with
  the input model by \cite{g07} (solid lines).  The contributions of
  unabsorbed, Compton-Thin and Compton-Thick AGN are plotted
  separately. }
\label{agn}
\end{figure}

\begin{figure}
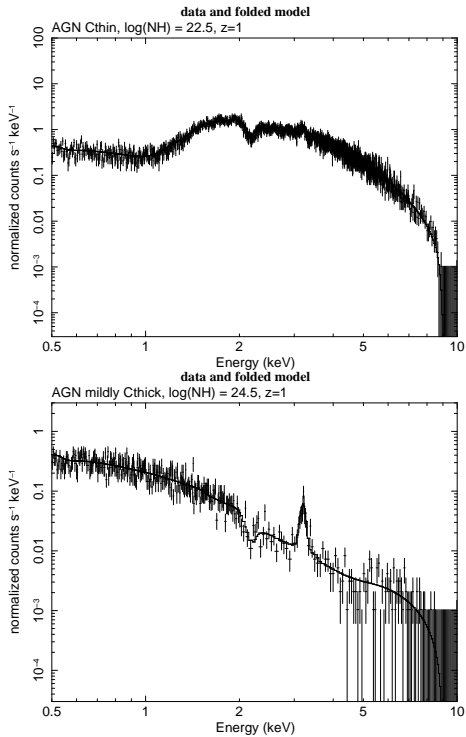

\centerline{\includegraphics[scale=0.25,angle=270]{AGN_cthin_22.5_z1.ps}} 
\centerline{\includegraphics[scale=0.25,angle=270]{AGN_mildcthick_model_z1.ps}}
\caption{Top panel: the spectrum of a typical compton-Thin AGN at z=1
  with intrinsic absorption $log(N_H)\sim 22.5$ cm$^{-2}$ and flux
  $F_S \sim 5 \times 10^{-14}$ erg s$^{-1}$ cm$^{-2}$ observed for
  13.2 ks with WFXT. Bottom panel: the spectrum of a mildly Compton
  Thick AGN at z=1 with nominal intrinsic absorption $log(N_H)\sim
  24.5$ cm$^{-2}$, soft flux $ \sim 5.4\times 10^{-14}$ erg s$^{-1}$
  cm$^{-2}$, observed for 13.2 ks with WFXT.  In both cases a soft
  scattered component in addition to the primary absorbed one has been
  considered.}
\label{spectra}
\end{figure}

\begin{figure}
\resizebox{\hsize}{!}{\includegraphics[clip=true]{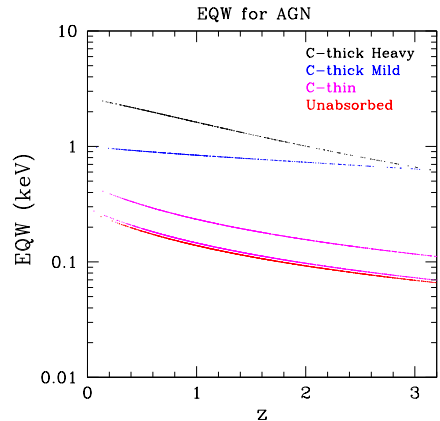}}
\caption{Observed-frame equivalent width of the 6.4 keV neutral Fe
  $K_\alpha$ line adopted in our AGN population for different AGN
  types, as a function of redshift.}
\label{eqw}
\end{figure}

In this work, we present only single--pointing images of extragalactic
fields.  The point source populations include four families of Active
Galactic Nuclei (Unabsorbed, Compton--Thin, Mildly Compton--Thick and
Heavily Compton--Thick), and normal and star forming galaxies,
consistently with the observed luminosity function and extrapolated to
high redshifts according to the \cite{g07} XRB synthesis model.  The
input logN-logS are shown in Figure \ref{agn}.  Two spectra of a
Compton Thin and a Compton Thick AGN, as observed by WFXT, are shown
in Figure \ref{spectra}.  Each AGN type is simulated with a neutral
Iron $K_\alpha$ line at 6.4 keV rest-frame with a typical equivalent
width as commonly observed in each source class, as shown in Figure
\ref{eqw}.  The presence of the Iron line allows one to measure the
redshift for a significant number of AGN, even though its equivalent
width is expected to vary significantly from source to source.  The
feature at 2 keV in the spectra is due to a dip in the effective area
(compare with Figure \ref{eff_area}).

In addition to AGN, another important class of point sources is
constituted by the star-forming galaxies.  As we learned from the
Chandra Deep Fields \citep{norman04,lehmer08}, star forming galaxies
are expected to dominate the number counts below fluxes $\sim
10^{-17}$ cgs.  WFXT will not reach fluxes lower than the limits
achieved in the Chandra Deep Fields, however it will be able to detect
thousands of star forming galaxies up to reshift one, providing an
unbiased view of the cosmic star formation history with an
unprecedented statistics (see contribution by P. Ranalli in this
volume).  We extract the galaxy catalog from the soft logN-logS
relations \citep{ran05}.  For all the galaxies we assumed an
intrinsically unabsorbed power law spectrum with photon index $\Gamma
= 2$.

\begin{figure}
\resizebox{\hsize}{!}{\includegraphics[clip=true]{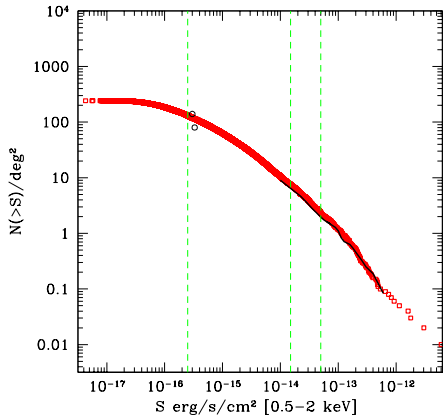}}
\caption{Soft band logN-logS for groups and clusters of galaxies from
  the input catalog of 100 square degrees (corresponding to 100 WFXT
  fields), extracted from the PS-like mass function model (red
  squares).  The black line refers to the ROSAT Deep Cluster Survey
  (RDCS) by \cite{rdcs}.  The two circles at low fluxes refer to the
  Chandra Deep Fields \citep{rosatiaa}.  Vertical dashed lines show
  the approximate flux limits for detection corresponding to the three
  WFXT surveys.}
\label{logNlogS_mf}
\end{figure}

\begin{figure}
\resizebox{\hsize}{!}{\includegraphics[clip=true]{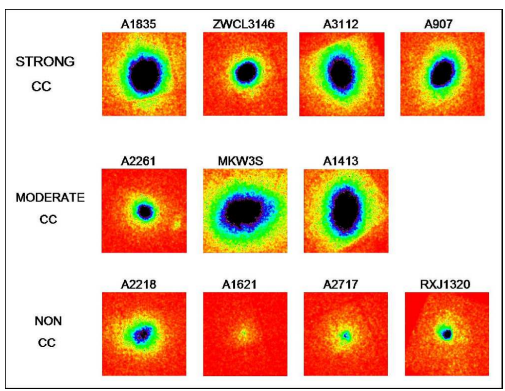}}
\caption{Cluster templates (original Chandra images) used in the
  simulations, representing three classes with different cool-core
  strength \citep{santos08}. }
\label{templates}
\end{figure}

Another important class of extragalactic sources are groups and
clusters of galaxies.  These sources are intrinsically extended since
the X-ray emission is due to the thermal bremsstrahlung in the hot
ICM.  We extracted a population of groups and clusters from the
\citet{ps74} mass function, tuning the cosmological parameters in a
$\Lambda$CDM universe in order to reproduce with reasonable accuracy
the existing constraints on the observed number counts (see Figure
\ref{logNlogS_mf}) and the observed luminosity \citep{rdcs} and
temperature functions \citep{henry09}.  Given their nature, X-ray
extended sources have a variety of different morphologies and
concentrations, which will be resolved in most cases thanks to the
WFXT angular resolution.  In order to render this aspect in our
simulations, we used real Chandra images (i.e., at very high
resolution) from a representative local sample of groups and clusters
in order to mimic the observed mix of flat and strongly peaked surface
brightness profiles (see Figure \ref{templates}).  This procedure has
been adapted from the cloning technique described in \cite{santos08}.
The presence of a cool--core may affect the detectability of the
cluster emission particularly at high redshift, and therefore it will
be an important aspect when quantifying the completeness of deep
cluster samples.  Clearly, some uncertainty is due to the evolution of
cool-cores, which is currently measured to be mild \citep{santos10}.
At present, we simply assume a fair mix of cool-core and non cool-core
clusters as observed locally.

\begin{figure*}
   \centering
\includegraphics[width=\textwidth]{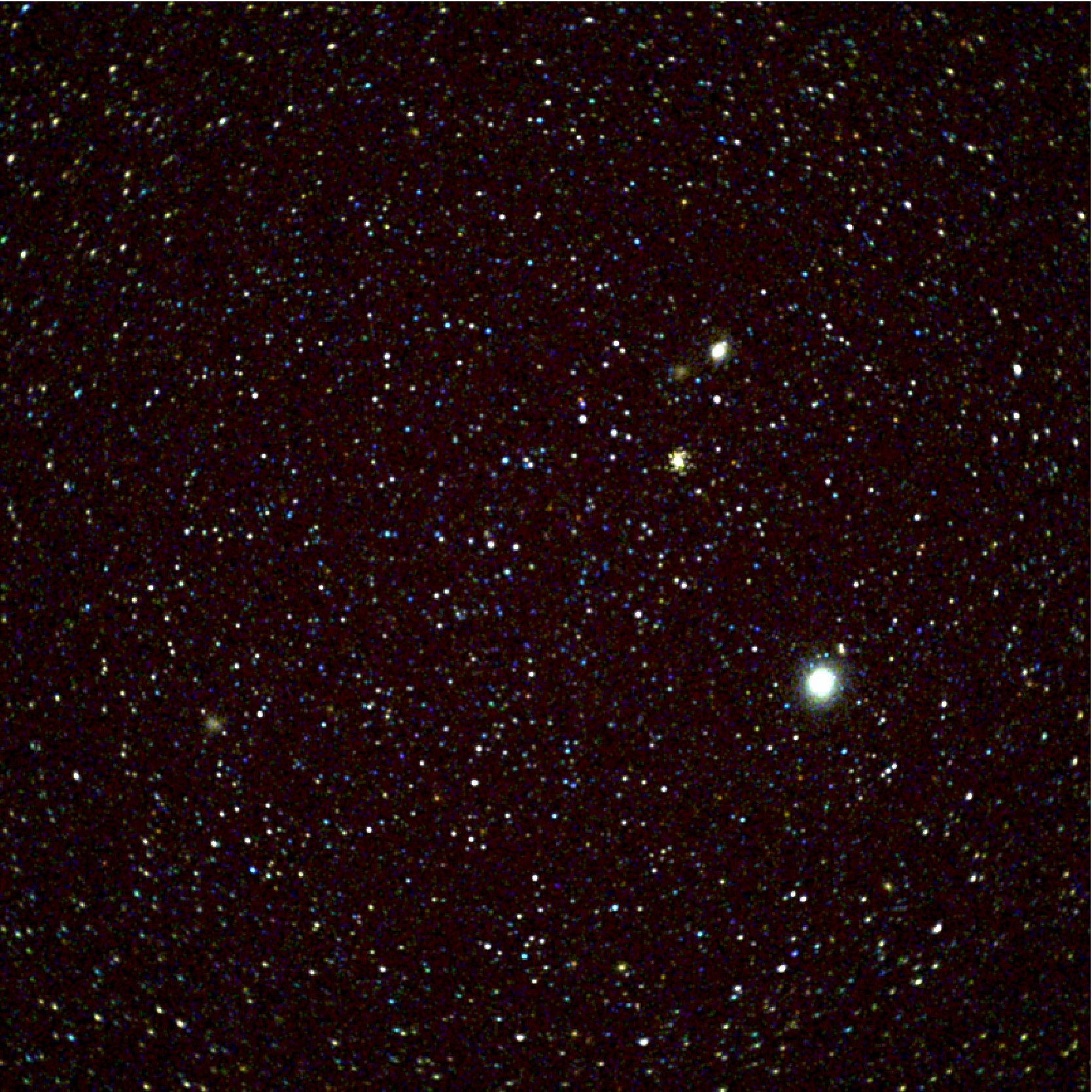}
\caption{A simulated WFXT extragalactic field with an exposure of 13.2
  ks (corresponding to one tile of the Medium survey). }
\label{wfxtfield}
\end{figure*}

The WFXT images are created in three bands (0.5-1 keV, 1-2 keV, and
2-7 keV) and combined to produce color images (see Figure
\ref{wfxtfield}).  We also produce the soft image (0.5-2 keV band)
which will be used for source detection, together with the 2-7 keV
hard band image.  Each image is given by the sum of the three modules.
The original pixel size is assumed to be 0.88 arcsec, while the final
images are resized by a factor of three (corresponding to a pixel of
2.64 arcsec) thus adequately sampling the PSF across the FOV.

We showed the efficiency of WFXT by comparing one tile of the medium
survey (corresponding to a single pointing of WFXT with a 13.2 ks
exposure) with the Chandra image of the COSMOS field \citep{cap09} in
the contribution by Rosati et al., this volume.  The striking result
is that WFXT is $\sim 150$ times faster in obtaining an image of the
same solid angle and same depth of Chandra COSMOS, and with a
resolution only a factor of two below that of the Chandra mosaic.
This direct comparison shows the tremendous survey efficiency of WFXT.

\section{Source detection with WFXT}

In principle, the large size of the WFXT images and the large number
of sources which will be detected would require dedicated software.
Indeed, we are planning to develope specific source detection
algorithms and source extraction procedures in the next future.  For
the purpose of this Paper, we use the algorithm {\sl wavdetect} which
has been developed for Chandra and it is part of the {\tt ciao}
software.  This procedure, despite not optimized for WFXT data, allows
us to perform a rapid test on the quality of our simulated images.

We run {\tt wavdetect} on the soft (0.5-2 keV) and hard (2-7 keV)
images of a tile of the Medium survey (corresponding to an exposure
time of 13.2 ks) with a standard set of parameters. The catalog of the
detected sources is then matched with the input sources.  In this way
we have a catalog of the matched sources, a list of spurious and a
list of undetected sources.

\begin{figure}
\resizebox{\hsize}{!}{\includegraphics[clip=true]{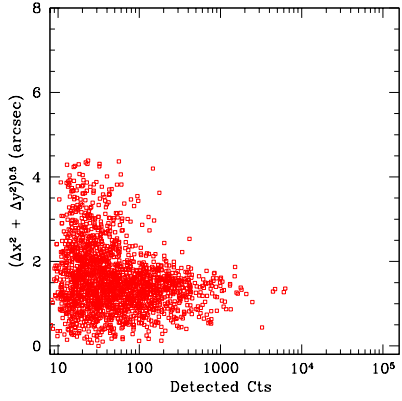}}
\caption{Position error in arcsec as a function of the net detected counts for
  point sources, from a simulated soft band image of the Medium survey
  (13.2 ks).}
\label{err}
\end{figure}

\begin{figure}
\resizebox{\hsize}{!}{\includegraphics[clip=true]{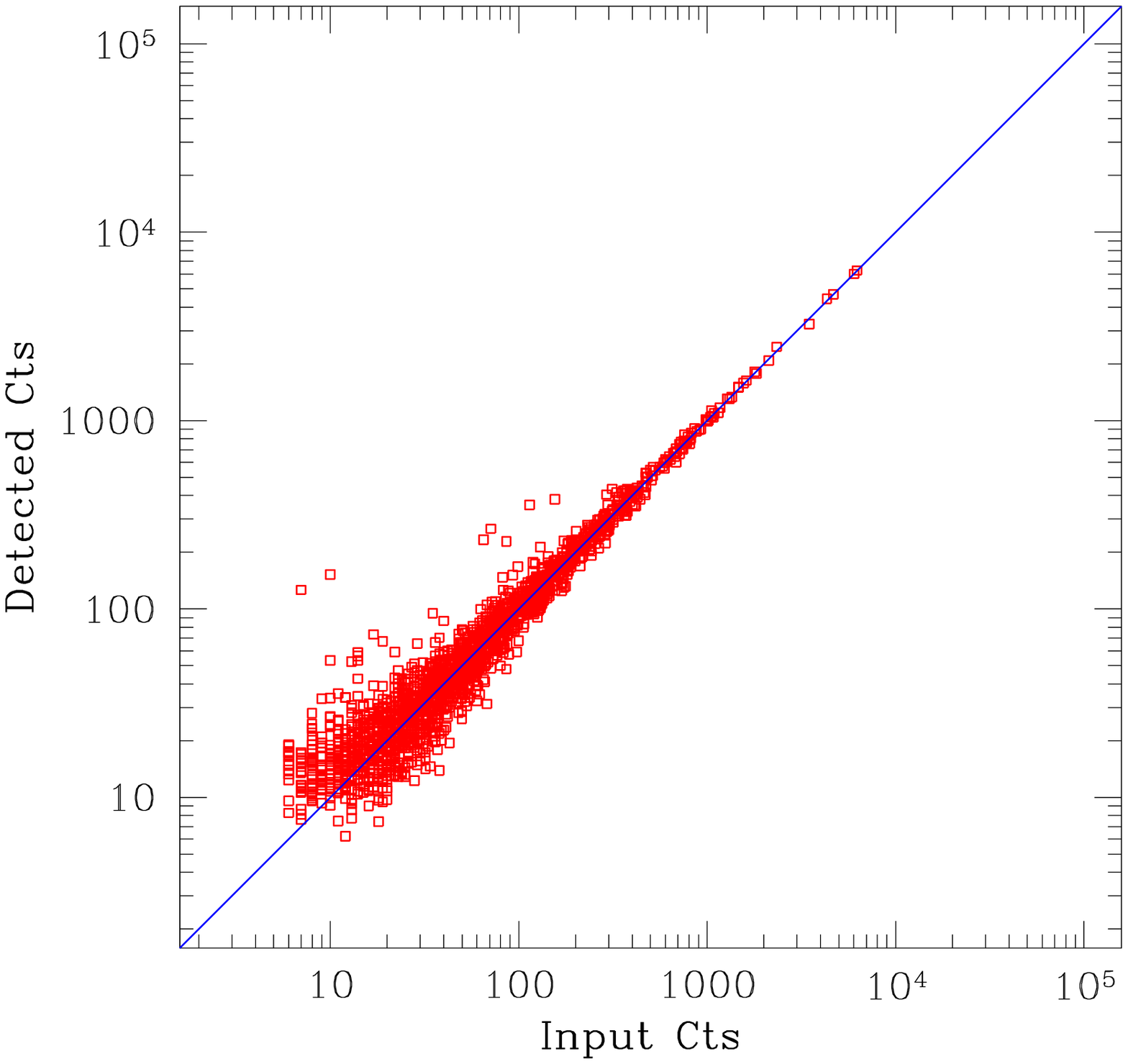}}
\caption{Recovered vs input counts in the soft band for point sources
  (from a simulated soft band image of the Medium survey).}
\label{counts_comparison}
\end{figure}

\begin{figure}
\resizebox{\hsize}{!}{\includegraphics[clip=true]{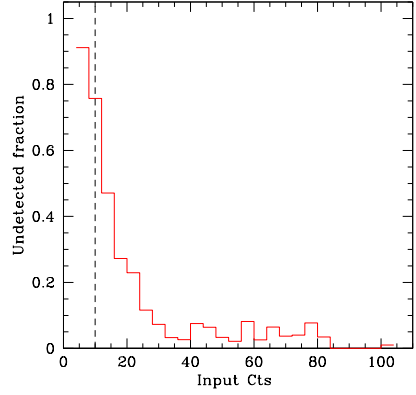}}
\caption{Histogram distribution of the undetected point sources in the
  soft band as a function of the input photons (from a simulated image
  of the Medium survey).
}
\label{undetected}
\end{figure}

Point sources are recovered with a position accuracy typically below 2
arcsec (see Figure \ref{err}).  Note that part of this error is due to
the asymmetry of the PSF, which is not corrected here, but it can be
accounted for when the PSF model will be included in the detection
algorithm.  The number of detected photons for each source is in very
good agreement with the input value within the poissonian error (see
Figure \ref{counts_comparison}).  The sources are recovered
efficiently down to 20 net photons, a value below which the number of
undetected sources grows rapidly (see Figure \ref{undetected}).  This
value provide a very conservative estimate of the flux limit in WFXT
images, and it can be considered constant with respect to the exposure
time, since point sources are very mildly affected by the diffuse
background.  For an image of the Medium survey, 20 net photons
correspond to a flux of $3.3 \times 10^{-16}$ erg s$^{-1}$ cm$^{-2}$
in the soft band.  At the same time, thanks to the low background, the
number of spurious sources is negligible above 20 counts.  We conclude
that the constant PSF of WFXT allows one to detect sources with an
almost flat sky coverage and negligible contamination down to very low
flux levels.  Eventually, it will be possible to decrease the
detection limit by a factor of two with a more sophisticated detection
algorithm.

\begin{figure}
\resizebox{\hsize}{!}{\includegraphics[clip=true]{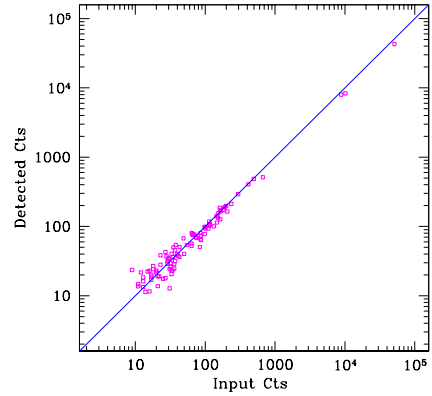}}
\caption{Recovered vs input counts in the soft band for extended
  sources (from a simulated soft band image of the Medium survey).}
\label{counts_comparison_ext}
\end{figure}

The detected photons are in agreement with the input value also for
extended sources (groups and clusters of galaxies) once a small offset
due to the lost emission at low surface brightness is accounted for.
However, at present, our schematic detection algorithm is not
efficient in characterizing extended sources.  Therefore, we set a
conservative detection limit by requiring a typical S/N ratio of 5
from aperture photometry.  To do this, we assume that for a typical
group or cluster at medium and large redshifts, roughly 80\% of the
flux is included within 30 arcsec, corresponding to an extraction area
of $A_{ext} = 2.18 \times 10^{-4}$ deg$^2$.  We adopt an average
conversion factor of $2.2 \times 10^{-13}$ erg s$^{-1}$ cm$^{-2}$ for
a typical ICM emission as in Table \ref{cf_sim_table}, and we use the
background values shown in Table \ref{bck_tot}.  We compute the S/N
ratio as follows:


\begin{equation}
BCK = BCK_{rate} \times  T_{exp} \times A_{ext} \, ,
\end{equation}

\begin{equation}
S/N = CTS_{net}/\sqrt{CTS+2 \, BCK} \, .
\end{equation}

Requiring $S/N > 5$, the condition on the minimum number of net
detected photons within the extraction regions is:

\begin{equation}
CTS_{det} = 12.5 \times (1 + \sqrt{1+0.32 \times BCK}) \, ,
\end{equation}

\noindent
corresponding to a flux of $F_{det} = CTS_{det} \times ECF/T_{exp} /
0.8$, where the factor of 0.8 accounts for the lost flux (see Table
\ref{det_ext}).

\begin{table}
\centering                          
\begin{tabular}{cccc}
 \hline\hline\\
Survey & $CTS_{ext}$  & $CTS_{tot}$  & $F_{lim}/10^{-15}$\\
\hline\hline\\
Wide &   39.4   & 49.3 & $5.42$ \\ 
Medium &  69  &  86.1 &  $1.44$ \\ 
Deep &    294   &  367 &  $0.2$ \\ 
 \hline\hline
\end{tabular}
\caption{Detection limits for extended sources: $CTS_{ext}$ is the
  minimum number of detected photons in the extraction regions,
  $CTS_{tot}$ is the detection limit in terms of total emitted
  photons, and $F_{lim}$ is the total flux limit.}
\label{det_ext}
\end{table}

\section{Redshift measure via X-ray spectroscopy of distant clusters} 

\begin{figure}
\centerline{\includegraphics[scale=0.36,angle=0]{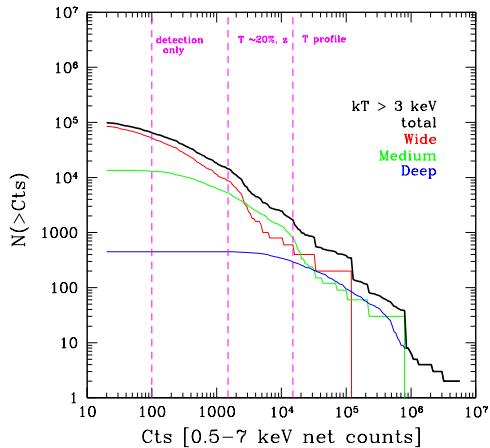}}
\caption{Number of clusters with $kT > 3$  keV for the three WFXT
  surveys as a function of the detected photons (0.5-7 keV band).
  Vertical dashed lines correspond roughly to the average detection
  limit (100 photons), to the limit for spectral analysis (1500 net
  photons) and the the limit for spatially resolved spectral analysis
  (15000 net photons).}
\label{cl_survey}
\end{figure}

Now we focus on the properties of the WFXT sample of X-ray clusters.
As we showed in this work, all the properties of the WFXT concur to
deliver a very high quality set of data: the large effective area
provides a high number of detected photons for sources down to low
fluxes; the angular resolution allows us to avoid source confusion and
remove the contribution of point sources from the diffuse emission of
clusters and groups; finally the large field of view allows one to
collect a large number of sources in a reasonable amount of time.  To
quantify the expected sample of clusters with $kT>3 $ keV from the
three WFXT surveys, we show in Figure \ref{cl_survey} the number of
extended sources as a function of the detected photons.

Despite the bulk of the detected photons will be in the soft band, the
relatively high effective area in the hard band (2-7 keV) will allow
us to measure the ICM temperature and detect the $K_\alpha$ Fe line at
any redshift whenever the equivalent width is $few \times 100$ eV.
For the first time, this opens the possibility of building a sample of
clusters with measured redshifts without recurring to time-consuming
optical follow-up work.

\begin{figure}
\resizebox{\hsize}{!}{\includegraphics[clip=true]{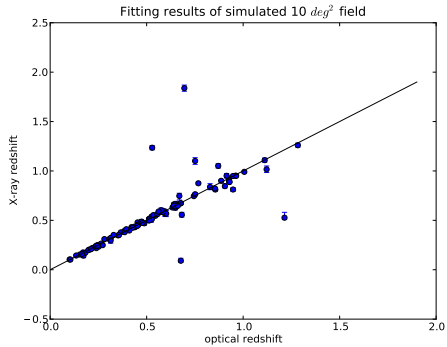}}
\caption{X-ray measured redshift vs optical (input) redshift from the
  spectral analysis of a mock simulation of 10 square degrees of the
  Medium survey (exposure time of 13.2 ks).  Vignetting effects are
  included (from Yu et al. 2010).}
\label{zx_vs_zopt}
\end{figure}

To explore this relevant science case, we performed spectral
simulations of a sample of groups and clusters extracted from the
\citet{ps74} mass function, whose temperatures and luminosities are
assigned according to the observed $M$-$T$ and $L$-$T$ relations.  The
X-ray spectra are analyzed with {\tt Xspec} with a {\tt mekal} model
where the redshift is left free to vary.  Even though the Fe
$K_\alpha$ line complex is ubiquitous in the ICM emission, the blind
search of the Fe line is a difficult task.  The background and the
poisson noise in the rapidly decreasing signal in the hard band, may
originate spurious lines which lead to catastrophic errors on the
measured X-ray redshift.  Another source of uncertainty is the
intrinsic Fe abundance, which is significantly varying particularly in
groups.  Our automatic procedure to find the X-ray redshift (Yu et
al. 2010, in preparation) shows that we will be able to detect the Fe
$K_\alpha$ line complex in any extended source detected with more than
800 total photons.  The number of catastrophic failures ($\Delta z
\geq 0.1$) is kept low, and the typical error on the redshift is
$\langle \Delta z \rangle \sim 0.022$.  This can be seen in Figure
\ref{zx_vs_zopt}.

\begin{figure}
\resizebox{\hsize}{!}{\includegraphics[clip=true]{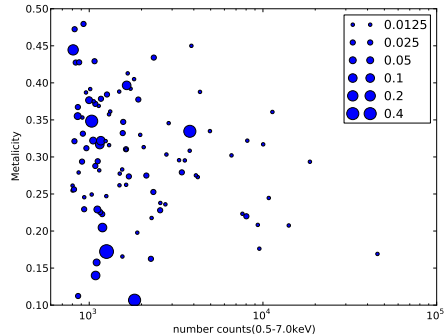}}
\caption{Input Fe abundance vs detected number counts for the mock
  simulation of 10 square degrees of the Medium survey.  The size of
  each dot is proportional to the error $\Delta z$ with respect to the
  input value (from Yu et al. 2010).}
\label{Z_vs_cts}
\end{figure}

The occurrence of catastrofic errors depends strongly on the total
detected photons, and mildly on the intrinsic Fe abundance.  As shown
in Figure \ref{Z_vs_cts}, where only clusters with more than 800
detected photons are considered, we can conservatively adopt a lower
limit of 1500 net detected photons above which we can rely on the
X-ray measured redshift.  We can straightforwardly compute that the
three WFXT planned surveys will provide a total of 15000-20000
clusters with $kT > 3 $ keV (as shown in Figure \ref{cl_survey}) with
measured X-ray redshift, a {\sl golden sample} that can be directly
used for precision cosmological tests with an unprecedented statistics
(see Borgani et al. this volume).
 
\section{Conclusions}

The Wide Field X-ray Telescope has been designed in order to be
optimized for surveys.  Its goal is not only to detect the largest
number of X-ray sources, but also to characterize them in order to
address several science cases.  The superiority of WFXT as an X-ray
survey machine, is due to the combination of angular resolution ($HEW
\sim 5$ arcsec), high effective area, and good sensitivity in the hard
band.  These properties open the discovery space of many scientific
cases that are not addressed by any other present of future X--ray
mission.

Thanks to the detailed imaging and spectral simulations presented in
this work, we have investigated, among the many scientific cases
within reach of WFXT, the construction of a cluster sample as large as
15000-20000 with redshift measured from the X-ray spectra, suitable
for high precision cosmological tests.  WFXT is therefore in the
position of achieving the maximum scientific impact thanks to the
high--quality characterization of the detected sources, of providing
an extremely large number of targets suitable for next-generation
X-ray missions like the International X-ray Observatory (IXO), with a
unique legacy value to be added to the next--generation wide field
surveys at other wavelengths.

The simulation tool presented in this work will provide an important
testbed to refine the design of the mission in keeping with the
scientific requirements.  In particular, the imaging and spectral
simulations will be used to develop specific detection algorithms and
to refine analysis procedures.  On the basis of the end-to-end
analysis presented here, we reinforce our idea that an X-ray mission
optimized for surveys like WFXT is necessary to bring X-ray astronomy
at the level of the optical, IR, submm and radio wavebands as foreseen
in the coming decade.


\begin{acknowledgements}
We acknowldge support under the ASI grant I/088/06/0 and the INFN PD51
grant.
\end{acknowledgements}

\bibliographystyle{aa}

\begin{thebibliography}{19}
\expandafter\ifx\csname natexlab\endcsname\relax\def\natexlab#1{#1}\fi

\bibitem[{{Burrows} {et~al.}(1992){Burrows}, {Burg}, \& {Giacconi}}]{bbg92}
{Burrows}, C.~J., {Burg}, R., \& {Giacconi}, R. 1992, \apj, 392, 760

\bibitem[{{Cappelluti} {et~al.}(2009){Cappelluti}, {Brusa}, {Hasinger},
  {Comastri}, {Zamorani}, {Finoguenov}, {Gilli}, {Puccetti}, {Miyaji},
  {Salvato}, {Vignali}, {Aldcroft}, {B{\"o}hringer}, {Brunner}, {Civano},
  {Elvis}, {Fiore}, {Fruscione}, {Griffiths}, {Guzzo}, {Iovino}, {Koekemoer},
  {Mainieri}, {Scoville}, {Shopbell}, {Silverman}, \& {Urry}}]{cap09}
{Cappelluti}, N., {Brusa}, M., {Hasinger}, G., {et~al.} 2009, \aap, 497, 635

\bibitem[{{Conconi} {et~al.}(2010){Conconi}, {Campana}, {Tagliaferri},
  {Pareschi}, {Citterio}, {Cotroneo}, {Proserpio}, \& {Civitani}}]{conc}
{Conconi}, P., {Campana}, S., {Tagliaferri}, G., {et~al.} 2010, \mnras, 405,
  877

\bibitem[{{Giacconi} {et~al.}(2009){Giacconi}, {Borgani}, {Rosati}, {Tozzi},
  {Gilli}, {Murray}, {Paolillo}, {Pareschi}, {Tagliaferri}, {Ptak},
  {Vikhlinin}, {Flanagan}, {Weisskopf}, {Bignamini}, {Donahue}, {Evrard},
  {Forman}, {Jones}, {Molendi}, {Santos}, \& {Voit}}]{giac09}
{Giacconi}, R., {Borgani}, S., {Rosati}, P., {et~al.} 2009, in ArXiv
  Astrophysics e-prints, Vol. 2010, astro2010: The Astronomy and Astrophysics
  Decadal Survey, 90

\bibitem[{{Gilli} {et~al.}(2007){Gilli}, {Comastri}, \& {Hasinger}}]{g07}
{Gilli}, R., {Comastri}, A., \& {Hasinger}, G. 2007, \aap, 463, 79

\bibitem[{{Henry} {et~al.}(2009){Henry}, {Evrard}, {Hoekstra}, {Babul}, \&
  {Mahdavi}}]{henry09}
{Henry}, J.~P., {Evrard}, A.~E., {Hoekstra}, H., {Babul}, A., \& {Mahdavi}, A.
  2009, \apj, 691, 1307

\bibitem[{{Lehmer} {et~al.}(2008){Lehmer}, {Brandt}, {Alexander}, {Bell},
  {Hornschemeier}, {McIntosh}, {Bauer}, {Gilli}, {Mainieri}, {Schneider},
  {Silverman}, {Steffen}, {Tozzi}, \& {Wolf}}]{lehmer08}
{Lehmer}, B.~D., {Brandt}, W.~N., {Alexander}, D.~M., {et~al.} 2008, \apj, 681,
  1163

\bibitem[{{Murray} {et~al.}(2009){Murray}, {Gilli}, {Tozzi}, {Paolillo},
  {Brandt}, {Tagliaferri}, {Vikhlinin}, {Bautz}, {Allen}, {Donahue},
  {Flanagan}, {Rosati}, {Borgani}, {Giacconi}, {Weisskopf}, {Ptak}, {Gezari},
  {Alexander}, {Pareschi}, {Forman}, {Jones}, \& {Hickox}}]{mur09}
{Murray}, S., {Gilli}, R., {Tozzi}, P., {et~al.} 2009, in ArXiv Astrophysics
  e-prints, Vol. 2010, astro2010: The Astronomy and Astrophysics Decadal
  Survey, 217

\bibitem[{{Murray} {et~al.}(2008){Murray}, {Norman}, {Ptak}, {Giacconi},
  {Weisskopf}, {Ramsey}, {Bautz}, {Vikhliniin}, {Brandt}, {Rosati}, {Weaver},
  {Allen}, \& {Flanagan}}]{mur08}
{Murray}, S., {Norman}, C., {Ptak}, A., {et~al.} 2008, in Society of
  Photo-Optical Instrumentation Engineers (SPIE) Conference Series, Vol. 7011,
  Society of Photo-Optical Instrumentation Engineers (SPIE) Conference Series

\bibitem[{{Norman} {et~al.}(2004){Norman}, {Ptak}, {Hornschemeier}, {Hasinger},
  {Bergeron}, {Comastri}, {Giacconi}, {Gilli}, {Glazebrook}, {Heckman},
  {Kewley}, {Ranalli}, {Rosati}, {Szokoly}, {Tozzi}, {Wang}, {Zheng}, \&
  {Zirm}}]{norman04}
{Norman}, C., {Ptak}, A., {Hornschemeier}, A., {et~al.} 2004, \apj, 607, 721

\bibitem[{{Predehl} {et~al.}(2010){Predehl}, {B{\"o}hringer}, {Brunner},
  {Brusa}, {Burwitz}, {Cappelluti}, {Churazov}, {Dennerl}, {Freyberg},
  {Friedrich}, {Hasinger}, {Kendziorra}, {Kreykenbohm}, {Schmid}, {Wilms},
  {Lamer}, {Meidinger}, {M{\"u}hlegger}, {Pavlinsky}, {Robrade}, {Santangelo},
  {Schmitt}, {Schwope}, {Steinmetz}, {Str{\"u}der}, {Sunyaev}, \&
  {Tenzer}}]{pr09}
{Predehl}, P., {B{\"o}hringer}, H., {Brunner}, H., {et~al.} 2010, in American
  Institute of Physics Conference Series, Vol. 1248, American Institute of
  Physics Conference Series, ed. {A.~Comastri, L.~Angelini, \& M.~Cappi},
  543--548

\bibitem[{{Press} \& {Schechter}(1974)}]{ps74}
{Press}, W.~H. \& {Schechter}, P. 1974, \apj, 187, 425

\bibitem[{{Ptak} {et~al.}(2009){Ptak}, {Feigelson}, {Chu}, {Kuntz}, {Zezas},
  {Snowden}, {de Martino}, {Trinchieri}, {Gabbiano}, {Forman}, {Tagliaferri},
  {Giacconi}, {Murray}, {Allen}, {Bautz}, {Borgani}, {Brandt}, {Campana},
  {Donahue}, {Flannagan}, {Gilli}, {Jones}, {Miller}, {Pareschi}, {Rosati},
  {Schneider}, {Tozzi}, \& {Vikhlinin}}]{ptak09}
{Ptak}, A., {Feigelson}, E., {Chu}, Y., {et~al.} 2009, in ArXiv Astrophysics
  e-prints, Vol. 2010, astro2010: The Astronomy and Astrophysics Decadal
  Survey, 240

\bibitem[{{Ranalli} {et~al.}(2005){Ranalli}, {Comastri}, \& {Setti}}]{ran05}
{Ranalli}, P., {Comastri}, A., \& {Setti}, G. 2005, \aap, 440, 23

\bibitem[{{Rosati} {et~al.}(2002){Rosati}, {Borgani}, \& {Norman}}]{rosatiaa}
{Rosati}, P., {Borgani}, S., \& {Norman}, C. 2002, \araa, 40, 539

\bibitem[{{Rosati} {et~al.}(1998){Rosati}, {della Ceca}, {Norman}, \&
  {Giacconi}}]{rdcs}
{Rosati}, P., {della Ceca}, R., {Norman}, C., \& {Giacconi}, R. 1998, \apjl,
  492, L21+

\bibitem[{{Santos} {et~al.}(2008){Santos}, {Rosati}, {Tozzi}, {B{\"o}hringer},
  {Ettori}, \& {Bignamini}}]{santos08}
{Santos}, J.~S., {Rosati}, P., {Tozzi}, P., {et~al.} 2008, \aap, 483, 35

\bibitem[{{Santos} {et~al.}(2010){Santos}, {Tozzi}, {Rosati}, \&
  {Boehringer}}]{santos10}
{Santos}, J.~S., {Tozzi}, P., {Rosati}, P., \& {Boehringer}, H. 2010, ArXiv
  e-prints

\bibitem[{{Vikhlinin} {et~al.}(2009){Vikhlinin}, {Murray}, {Gilli}, {Tozzi},
  {Paolillo}, \& {et al.}}]{vik09}
{Vikhlinin}, A., {Murray}, S., {Gilli}, R., {et~al.} 2009, in Astronomy, Vol.
  2010, AGB Stars and Related Phenomenastro2010: The Astronomy and Astrophysics
  Decadal Survey, 305

\end{thebibliography}

\end{document}